\documentclass[aps,prb,twocolumn,showpacs]{revtex4}
\usepackage{tabularx,graphicx}
\usepackage{epsfig}
\newcommand{\cg}{\chi^C}
\newcommand{\cgq}{\chi^C({\bf \vec{q}},\omega)}
\newcommand{\csi}{\chi^G}
\newcommand{\csiq}{\chi^G({\bf \vec{q}},\omega)}
\newcommand{\nusi}{\nu^G}
\newcommand{\nug}{\nu^C}
\newcommand{\wq}{\omega_{\bf\vec{q}}}
\newcommand{\e}{\epsilon}

\newcommand{\Dr}{\Delta_{r}}

\newcommand{\om}{\omega}

\newcommand{\wco}{\omega_{co}}
\newcommand{\wo}{\omega_{0}}

\newcommand{\Dox}{\Delta^x_0}

\newcommand{\Doz}{\Delta^z_0}

\newcommand{\vf}{v_{\rm F}}
\newcommand{\kf}{k_{\rm F}}

\begin{document}
\title{Dissipation in graphene and nanotube resonators.}

\author{C. Seo\'anez}
\affiliation{Instituto de  Ciencia de Materiales de Madrid, CSIC,
 Cantoblanco E28049 Madrid, Spain}

\author{F. Guinea}
\affiliation{Instituto de  Ciencia de Materiales de Madrid, CSIC,
 Cantoblanco E28049 Madrid, Spain}

\author{A.~H. Castro Neto}
\affiliation{Department of Physics, Boston University, 590
Commonwealth Avenue, Boston, MA 02215, USA}


\begin{abstract}
Different damping mechanisms in graphene nanoresonators are studied:
charges in the substrate, ohmic losses in the substrate and the
graphene sheet, breaking and healing of surface bonds (Velcro
effect), two level systems, attachment losses, and thermoelastic
losses. We find that, for realistic structures and contrary to
semiconductor resonators, dissipation is dominated by ohmic losses
in the graphene layer and metallic gate. An extension of this study
to carbon nanotube-based resonators is presented.

\end{abstract}
 \maketitle
\section{Introduction.}
Nano-electro-mechanical devices\cite{C02,B04,ER05} (NEMS) have
attracted a great deal of attention, as they are a problem of
fundamental interest, and also because of their potential
applications.

Recently, NEMs made from graphene sheets have been
studied\cite{Betal07}, following work on NEMs based on carbon
nanotubes\cite{Setal04,S06}. These devices show unique
characteristics, as graphene sheet stacks have a high elastic
modulus and very small total mass thanks to the low number of atomic
planes (sometimes just even one) composing the bridge or cantilever,
allowing for higher resonating frequencies than other materials of
similar dimensions and increased potential sensitivity.

Graphene itself has attracted a great deal of attention\cite{GN07},
because of its unique features. Graphene samples can be made one
carbon layer thick, and doped by an external electric field. The
lattice dynamics of these thin samples have not been studied in
detail yet. Two dimensional systems have, in addition to acoustic
modes, transverse flexural modes\cite{WR04}, which are the ones
explored in the experiments considered in this paper. The quadratic
dispersion of these modes lead to a constant density of states at
low energy. Experimental observations show that free standing
graphene is not flat\cite{Metal07,ICCFW07,Setal07}, but corrugated.
These ripples imply the existence of flexural deformations, and can
lead to charge inhomogeneities in single layer graphene at low
dopings\cite{Metal07b,CF07}. Most graphene samples stand on SiO$_2$
substrates\cite{Netal04}, and the interaction between the graphene
layers and the substrate is not well known. The experiments
discussed in this paper can provide information on this issue.

The potential sensitivity of a resonator-based detector may be in
practice strongly reduced by dissipative processes affecting the
vibrational mode used for detection, due to the associated widening
of the resonance, which masks the frequency shifts used to determine
the presence of external species adsorbed or close to the detector.
Hence, it is of fundamental importance to gain knowledge about those
damping mechanisms, to establish their relative importance and
dependence on resonator parameters (dimensions, elastic constants,
temperature, etc), which may help to optimize performances and
determine where should efforts be put, not only to use them as
detectors, but also as tools for the study of more fundamental
questions like the quantum to classical crossover with increasing
system sizes \cite{B04,ZGBM05}.

In common resonating structures made of semiconductors the
prevailing dissipative mechanism with decreasing size and
temperature are surface-related losses: the presence of the
imperfect surface, with its roughness, structural defects,
impurities and dangling bonds, can be modeled by a distribution of
effective two-level systems which couple to the vibrational
eigenmodes of the device \cite{SGN07,YOE02,C07}. Many other
processes contribute to a lower extent to the damping of vibrations
in these devices. Some of them are common to all experimental
setups, like attachment losses \cite{JI68,PJ04} or thermoelastic
damping \cite{Z38,LR00,U06}. Others depend on the actuation scheme:
For example, in the magnetomotive actuation scheme\cite{Metal02}, a
layer of metal is deposited on top of the vibrating semiconducting
structure to control its motion with the Lorentz force actuating on
the electron current that passes through the top layer in presence
of an applied magnetic field. This metallic layer increases
dissipation in two different ways: i) Increasing the local
temperature due to electron-phonon interactions, thus "feeding"
other mechanisms whose effect tends to grow with T, ii) Absorbing
energy through the excitation of electron-hole pairs in the metallic
layer due to the presence of fixed charges in the substrate
supporting the oscillating structure, which create a potential on
the electrons moving within the mobile structure that is
time-dependent, as perceived by the latter.

This last mechanism has been ignored until now in the literature,
perhaps due to the small amount of fixed charges in typical
substrates like single-crystal Si or GaAs. But in the case of
graphene or carbon-nanotube based resonators it must be considered,
as it plays a much more significant role. This is due to the fact
that graphene is conducting, and in some actuation setups, the
control over the graphene layer's motion is through the
establishment of a capacitive coupling between two charged layers,
namely the oscillating graphene and a doped Si backgate. The number
of carriers in both can be controlled by an external
gate\cite{Netal04}. The coupling between the charges of both layers,
apart from enabling the control of the resonator's motion, causes
energy losses which will dominate at high temperatures. There will
be also fixed charges in the supporting structure, mainly in the
SiO$_2$ layer located between the graphene and the doped Si
backgate, absorbing energy too from the resonators motion. In this
paper we give will give full account of these processes.

In the following, we start analyzing different dissipative processes
which may be present in devices based on graphene. Our calculations
should give reasonable order of magnitude estimates of the strengths
of these mechanisms. In Sections II and III we model the absorption
of mechanical energy due to the charge present in the oscillating
graphene sheet, the SiO$_2$ substrate and the Si backgate. In
section IV we discuss the role as attenuation source played in these
peculiar devices by the breaking and healing of bonds gluing the
graphene sheet to the SiO$_2$ substrate, a possibility also missing
in current semiconducting resonators. Whereas for the latter prevail
surface-related losses, we show in Section V that for graphene
resonators this friction mechanism is highly suppressed thanks to
their high degree of crystallinity. For completeness, we apply
previous results of other works to estimate the effect of two more
friction sources present in all setups, namely attachment losses and
thermoelastic losses, in Section VI. Once these mechanisms have been
studied, an extension of the results to carbon nanotube resonators
is presented in Section VII.

\begin{figure}
\begin{center}
\includegraphics[width=7cm]{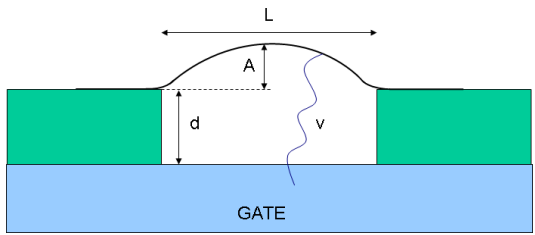}\\
\caption[fig]{\label{sketch} (Color online) Sketch of the system
considered in the text. $v$ represents the Coulomb interaction
between the charged graphene layer and the Si gate.}
\end{center}
\end{figure}

To make numerical estimates we focus on the devices studied
in\cite{Betal07}. The parameters which characterize the average
oscillator studied there are given in Table[\ref{table_parameters}].
A sketch of the system is shown in Fig.[\ref{sketch}].

\begin{table}
 \begin{tabular}[c]{||c|c||}
  \hline \hline
  \multicolumn{2}{||c||}{System properties}  \\ \hline
  Dimensions & \\ Thickness $t$ & $10\cdot10^{-9}$ m\\
  Width $w$ & $10^{-6}$ m \\
  Length $L$ & $10^{-6}$ m \\
   Height above substrate
$d$ & $300\cdot10^{-9}$ m \\ \hline Frequency  $f_0$
   & 100 MHz \\
Amplitude $A$ &0.5 nm \\ \hline Carrier density $\rho_C$ &$10^{12}
{\rm cm}^{-2}$ \\ \hline \hline \multicolumn{2}{||c||}{Properties of
graphite} \\ \hline Mass density $\rho_M^C$  & 2200 kg/m$^{3}$ \\
\hline
  Elastic constants & \\
$E$ & $10^{12}$ Pa \\
  $\nu$ & 0.16 \\ \hline
Debye temperature $\theta_{D}$ & $\sim
  570\,K$ \\ \hline
 Specific heat $C_p$ & 700 J / Kg. K \\ \hline
Thermal conductivity $\kappa$ & 390 W / m . K \\ \hline \hline

\end{tabular}
\caption{Parameters used in the calculations presented in the main
text,
  adapted to the systems studied in\protect{\cite{Betal07}}. Bulk data taken
  from\protect{\cite{P93}}.}
\label{table_parameters}
\end{table}

\section{Static charges at the amorphous SiO$_2$ substrate.}
The graphene sheet can couple electrostatically to static charges.
These charges give rise to a time dependent potential acting on the
electrons of the vibrating graphene. The energy is dissipated by
creating electron-hole excitations in the graphene layer. Static
charges have been proposed as a source of scattering by the carriers
in the graphene\cite{NM07,HAS06}.

The time-dependent component of the unscreened potential induced by
a charge separated by a distance $d$ in the vertical direction from
the graphene layer, acting on an electron at position ${\bf
\vec{r}}$ in the graphene layer is given, approximately, by:
\begin{equation}
V ( {\bf \vec{r}} , t ) \approx \frac{e^2 d A e^{i \omega_0 t}}{( |
{\bf  \vec{r}} |^2 + d^2 )^{3/2}} \label{potential_charge}
\end{equation}
where $A$ is the amplitude of the flexural mode, and $\omega_0$ its
frequency. The dissipation depends on the screening by the graphene
layer\cite{WSSG06}. A calculation of the damping is given in
Appendix \ref{app_charge}. We find that, for a single graphene
layer, a single charge gives a contribution to the inverse quality
factor of:
\begin{equation}
 Q^{-1} \approx \left\{ \begin{array}{lr} \frac{1}{\kf d}\frac{2\hbar}{M\omega_0d^2}
     &\kf d \gg 1 \\ \frac{2\hbar\omega_0^3 d^2}{M\vf^4} & \kf d \ll 1
    \end{array} \right.
\label{charge}
\end{equation}
where $M$ is the mass of the oscillating sheet, $\kf = \pi
\sqrt{\rho_C}$, and $\rho_C$ is the density of carriers in the
graphene sheet. Typical values of this quantity are in the range
$\rho_C \sim 10^{12} {\rm cm}^{-2}$, so that $\kf d \sim 10^2 - 10^3
\gg 1$. Eq.(\ref{charge}) can be generalized to a graphene sheet
with $N$ layers:
\begin{equation}
Q^{-1} \sim \frac{1}{\sqrt{N \rho} d} \frac{2\hbar}{M\omega_0d^2}
\label{charge_N}
\end{equation}
The suppression with the number of layers is due to the increased
screening in this system.

The total contribution to the inverse quality factor is obtained by
multiplying eqs.(\ref{charge}) or (\ref{charge_N}) by the total
number of charges $N_{\rm ch}$. An upper bound to the density of
local charges, deduced from some models for the electric
conductivity of graphene\cite{NM07,HAS06}, is $\rho_{\rm ch} \sim
10^{12} {\rm cm}^{-2}$. Using the parameters in
Table[\ref{table_parameters}], we find $N_{\rm ch} \sim 10^4$ and
$Q^{-1} \sim 10^{-11}$ at low temperatures.

This mechanism leads to ohmic dissipation, as the energy is
dissipated into electron-hole pairs in the metallic graphene layer.
Hence, the temperature dependence of this mechanism is given by
$Q^{-1} ( T ) \sim Q^{-1} ( 0 ) \times ( kT / \hbar\wo )$ , and
$Q^{-1} \sim 10^{-6}$ at 300 K.

\section{Ohmic losses at the graphene sheet and the metallic gate.}
The electrons in the vibrating graphene layer induce a time
dependent potential on the metallic gate which is sometimes part of
the experimental setup. The energy is transferred to electron-hole
pairs created at the gate or at the graphene layer. These processes
contribute to the energy loss and decoherence of electrons in
metallic conductors near gates\cite{GJS04,G05}.

The coupling between charge fluctuations in the two metallic systems
is due to long range electrostatic interactions. The corresponding
hamiltonian is
\begin{eqnarray}\label{Hwithint}
  \nonumber H &=& \frac{1}{2}\Bigl\{\int_Cv_{scr}(z,{\bf
    \vec{r}},t)\rho^C(z,{\bf \vec{r}},t) +\int_Gv_{scr}(0,{\bf
    \vec{r}}',t)\rho^G({\bf \vec{r}}')\Bigl\}\\
   && +\int_C\frac{1}{2\rho_M
    tw}\Pi^2+\frac{1}{2}\frac{Et^3w}{12}\Bigl[ \Bigl(\frac{\partial^2 \phi}
{\partial x^2}\Bigr)^2 +\Bigl(\frac{\partial^2 \phi}
{\partial y^2}\Bigr)^2\Bigr]
\end{eqnarray}
where the indices $G$ and $C$ stand for the gate and graphene layer,
respectively. $\rho_M$ is the mass density of the graphene sheet,
and $t,w,E$ its thickness, width and Young modulus, whereas
$\phi({\bf\vec{r}},t)$ represents the vibrating amplitude field of
bending modes and $\Pi=\partial\textsl{L} / \partial\dot{\phi}$ is
its conjugate momentum ($\textsl{L}$ is the Lagrangian). The
self-consistent screened potentials $v_{scr}(z,{\bf \vec{r}},t)$,
$v_{scr}(0,{\bf \vec{r}},t)$ are calculated as a function of the
bare potentials $v_0(z,{\bf \vec{r}},t)$, $v_0(0,{\bf \vec{r}},t)$
in Appendix \ref{metallic_layers}.

As in the case of eq.(\ref{potential_charge}), the time-dependent
part of the bare potentials couples the electronic degrees of
freedom and the mechanical ones through the charge $\rho({\bf
\vec{r}})$ and amplitude of the vibrational mode, $A_{\bf \vec{q}}$,
and would give rise to a term in the quantized hamiltonian of the
form
\begin{equation}\label{intelectrmech}
    H_{int}\propto \rho({\bf \vec{r}})A_{\bf \vec{q}}\propto(b^\dagger_{{\bf
    \vec{q}}}+b_{{\bf
    \vec{q}}})\sum_{{\bf \vec{k}},{\bf \vec{k}'}}[c^\dagger_{{\bf \vec{k}}+{\bf
    \vec{k}'}}c_{{\bf \vec{k}}}+\text{h.c}]
\end{equation}
where $A_{\bf \vec{q}}$ and $\rho({\bf \vec{r}})$ have been
expressed in terms of creation and annihilation operators of phonons
$({\bf \vec{q}},\wq)$ and electrons of a 2D Fermi gas, respectively.

But a realistic model requires taking into account the screening of
the potential associated to these charge fluctuations. In terms of
the screened potentials, the induced broadening of the mode $({\bf
\vec{q}},\wq)$ of the graphene layer can be written, using Fermi's
Golden Rule, as \cite{GJS04}
\begin{eqnarray}\label{gamma_gate}
  \nonumber \Gamma(\wq) =&& \sum_{\alpha = {\rm G,C}} \int d^3
  {\bf \vec{r}} \int d^3{\bf\vec{r}}' \Bigl\{ \rm{Re}V_{\rm  scr}^\alpha ( {\bf \vec{r}} , \wq )\times    \\
   && \times \rm{Re} V_{\rm scr}^\alpha ( {\bf \vec{r}}' , \wq )\times \rm{Im} \chi^\alpha [{\bf \vec{r}}- {\bf \vec{r}}' , \wq ]\Bigr\}
\end{eqnarray}
The static screening properties, $\lim_{{\bf \vec{q}} \rightarrow 0}
{\rm Re} \chi^{\alpha}  ( {\bf \vec{q}} , 0 )$,  of the graphene
layer and the gate are determined by their electronic
compressibilities, $\nug$ and $\nusi$ respectively. We will assume
that the distance between the graphene and the gate is much larger
than the electronic elastic mean free path in either material, so
that their polarizability is well approximated by:
\begin{equation}
\chi^\alpha ( {\bf \vec{q}} , \omega ) \approx \frac{\nu^\alpha
D^\alpha |
  {\bf \vec{q}} |^2}{D^\alpha | {\bf \vec{q}} |^2 + i \omega}
\end{equation}
where $D^\alpha = \vf^\alpha l^\alpha$ is the diffusion constant,
and $l^\alpha$ is the elastic mean free path. The two dimensional
conductivity is $g^\alpha = \kf^\alpha l^\alpha$.

We assume the gate to be quasi two dimensional. This approximation
is justified when the distance between the gate and the graphene
layer is much larger than the width of the gate. In this situation,
the broadening of the mode, eq.(\ref{gamma_gate}), can be expressed
as
\begin{equation}\label{gammamode2}
    \Gamma(\wq)\approx\int d^2{\bf \vec{k}}|v_{scr}(d,{\bf \vec{k}},\wq)|^2\rm{Im}\cg+|v_{scr}(0,{\bf \vec{k}},\wq)|^2\rm{Im}\csi
\end{equation}

The screened potentials for a graphene layer oscillating in an
eigenmode $({\bf \vec{q}},\wq)$ of amplitude $A_{\bf \vec{q}}$, have
in a first approximation only one momentum component, $v_{scr}({\bf
\vec{k}},\wq)=v_{scr}({\bf \vec{q}},\wq)\delta({\bf \vec{k}}-{\bf
\vec{q}})$, and these components are (see Appendix
\ref{metallic_layers})
\begin{equation}\label{vscr0}
    \left\{
      \begin{array}{l}
         v_{scr}(d,{\bf \vec{q}},\wq)=\frac{q\Bigl[\cg\Bigl(e^{qd}+e^{-qd}\Bigr)-2\csi e^{qd}\Bigr]\rho_CA_{\bf \vec{q}}e^{-qd}}
          {2\cg\csi\Bigl(1- e^{-2qd}\Bigr)} \\
         v_{scr}(0,{\bf \vec{q}},\wq)=\frac{|{\bf \vec{q}}|\Bigl[-\nug\Bigl(e^{2qd}+1\Bigr)+2\nusi \Bigr]\rho_0A_{\bf \vec{q}}e^{-qd}}
         {2\nug\nusi\Bigl(1- e^{-2qd}\Bigr)}
      \end{array}
    \right.\,\,
\end{equation}
where $q=|{\bf \vec{q}}|$ and $\rho_C$ is the charge density in the
graphene layer. The results for $\Gamma(\wq)$ and $Q^{-1}(\wq)$ can
be formulated in terms of the total charge in the graphene layer,
$Q_C = \int d^2 {\bf \vec{r}} \rho_C \approx L \times w \times
\rho_C$.

In the limit of short separation between the layers, $d\ll L$, which
is the situation present in current experimental setups, one has
\begin{equation}\label{gammamode5}
    \Gamma(\wo)\approx\frac{\wo A^2 Q_C^2}{4d^2}\Bigl(\frac{1}{\nug D^C}+\Bigl(\frac{\nusi}{\nug}\Bigr)^2\frac{1}{\nusi D^G}\Bigr)
\end{equation}
The limit $D|{\bf\vec{q}}|^2\gg\om$ for the imaginary part of the
susceptibility of a dirty metal,
$\rm{Im}\chi({\bf\vec{q}},\om)\approx\om\nu/D|{\bf\vec{q}}|^2$, has
been used. The first term in the summation describes losses at the
graphene sheet, and the second at the gate. The associated inverse
quality factor, according to eq.(\ref{Qfactor1a}), is given by
\begin{equation}\label{gammamode6}
    Q^{-1}(\wo)\approx\frac{\hbar Q_C^2}{2M\wo d^2}\Bigl(\frac{1}{\nug D^C}+\Bigl(\frac{\nusi}{\nug}\Bigr)^2\frac{1}{\nusi D^G}\Bigr)
\end{equation}
To make numerical estimates, we use the parameters in
Table[\ref{table_parameters}], with $\nu^C(E)= E/2 \pi \hbar^2
v_F^2$, $v_F\approx10^6$ m/s for a single layer of graphene, and
$\nu^C(E)= ( N \gamma ) / \vf^2$ for a stack of $N$
layers\cite{GNP06}. Carriers in graphene stacks have large
mobilities\cite{Netal04}, and we take $D^C\nu^C\approx10^3$. Typical
charge densities for the graphene layer are $\rho_C \sim 10^{12}
{\rm cm}^{-2}$, leading to a total charge $Q_C \sim 10^4$. For these
parameters, the contribution of the graphene sheet is $Q^{-1} \sim
10^{-8}$. The relative contribution from the gate depends on the
distance to the graphene sheet. For a Si layer with $D^G\nu^G\approx
10^3$ and at short distances, the contribution to the damping from
the gate is of the same order as that of the graphene sheet.

Damping is associated to the creation of e-h pairs in a metal, which
implies that this mechanism is ohmic. The inverse quality factor
should increase linearly with temperature, leading to $Q^{-1} \sim
10^{-2}$ at 300 K.

\section{Breaking and healing of surface bonds: Velcro effect.}
In the fabrication process of the device, the graphene flake is
deposited on the SiO$_2$ substrate, and becomes linked to it through
hydrogen bonds created by the silanol groups (SiOH) present at the
substrate's surface. When the flake is set into motion, some of this
bonds may repeatedly break and heal (the Velcro effect\cite{VZ07}),
causing dissipation of the energy stored in the vibration. Numerical
estimates are difficult to make, but nevertheless two qualitative
arguments showing that its role in the damping is probably
negligible can be presented:

i) This mechanism is expected to be
temperature independent, in contrast with the strong decrease of
friction observed as temperature is lowered \cite{Betal07}.

ii) The elastic energy stored in a typical graphene oscillator of
lateral dimensions $w\sim1\mu$m is about 10eV, when the amplitude is
$\sim 1$nm. This means about $\sim 10^{-5}$eV per nm$^2$. On the
other hand, the energy per hydrogen bond is about $10^{-1}$eV, and
typical radical densities at SiO$_2$ surfaces are \cite{DPX98} $\sim
1 {\rm nm}^{-2}$. Hence the elastic
  energy available on average for each hydrogen bond is much less than the energy stored in the
  bond. Only rare fluctuations, where a significant amount of energy is
  concentrated in a small area will be able to break bonds, and to induce
  energy dissipation. Note, however, that this argument ceases to be valid
  for very large amplitudes $\gtrsim 30$nm. For higher amplitudes, this mechanism
  can induce significant losses.

\section{Dissipation due to two-level systems.} This is the
typical mechanism for the damping of sound waves in insulating
amorphous materials\cite{AHV72,E98,ERK04}.  An atom or a few atoms
can have two nearly degenerate configurations. A vibration modifies
the energy difference between these situations. This mechanism leads
to the damping of acoustic phonons in amorphous SiO$_2$\cite{E98}.
It is also expected to dominate friction in many NEMs\cite{Metal02}.
We expect the graphene sheet to show a high degree of crystallinity,
and we will only consider two-level systems (TLSs) in the rest of
the structure.

The TLSs can only dissipate energy if they are coupled to the
vibrating graphene sheet. A possible mechanism is the existence of
charge impurities associated to these defects (fluctuating charges),
which are electrostatically coupled to the conducting electrons in
the graphene.

We expect this mechanism to be less effective in the device
considered here than in NEMs made of semiconducting materials, as
now the TLSs reside in the SiO$_2$ substrate, not in the vibrating
structure. The coupling, arising from long range forces, will be in
comparison accordingly suppressed, by a factor of order $( a / d
)^n$, where $a$ is a length comparable to the interatomic
separation, and $n$ describes the decay of the coupling ($n=1$ for
the Coulomb potential between charged systems).

The temperature dependence of the contribution of TLSs to $Q^{-1}$
is determined by the density of states of the modes coupled to the
TLSs and the distribution of TLSs in terms of their paramenters
(tunneling amplitude $\Dox$ and bias $\Doz$)\cite{SGN07}.The
hamiltonian describing the coupling of the effective TLS's and the
oscillating graphene sheet is given by \cite{SGN07}
\begin{equation}\label{Ham_TLS}
    H=\e\,\sigma_x+\gamma\frac{\Dox}{\e}\sigma_z\sum_{\textbf{k}}\lambda_{\textbf{k}}(b_{\textbf{k}}+b_{\textbf{-k}}^{\dagger})+
\sum_{\textbf{k}}\hbar\om_{\textbf{k}}(b_{\textbf{k}}^{\dagger}b_{\textbf{k}}+\text{h.c.})
\end{equation}
where $\e=\sqrt{(\Dox)^2+(\Doz)^2}$, $\gamma$ is the coupling
constant, which will be strongly suppressed in these devices as
compared to attenuation of acoustic waves in amorphous materials,
$\gamma\sim1\text{eV}\times( a / d )^n$, $b_{\textbf{k}}^{\dagger}$
represent the phonon creation operators associated to the different
vibrational modes of a sheet, and
$\sum_{\textbf{k}}\lambda_{\textbf{k}}(b_{\textbf{k}}+b_{\textbf{-k}}^{\dagger})$
represents the coupling to the strain tensor $u_{ik}$. There are two
types, compression modes (longitudinal waves) and bending modes. The
damping is due to the initial transfer of energy from the
vibrational mode studied by the experimentalists to the TLSs, which
in a second step transfer this energy to the rest of the modes. The
properties of the spin-boson model, eq.(\ref{Ham_TLS}), are fully
determined by the power-law $s$ of the spectral
function\cite{Letal87}, $ J ( \omega ) \equiv \sum_k
\left|\gamma\lambda_k\Dox /\e \right|^2 \delta ( \omega - \omega_k
)\sim \alpha\wco^{1-s}\om^s $, where $\omega_k$ is the frequency of
mode $k$, $\alpha$ is an adimensional constant and $\wco$ is the
upper cutoff of the phonon bath. For this system, compression modes
gives rise to a superohmic, $s=2$, bath, while the bending modes
constitute an ohmic bath, $s=1$, and thus will prevail as a source
of dissipation at low temperatures\cite{Letal87}. We will therefore
restrict our analysis to the dissipation caused by the ohmic
component of the vibrational spectrum.\\
Applying the method in\cite{SGN07} to the 2D bending modes of the
graphene sheet, one arrives at $J(\om)=\alpha\om$, with
\begin{equation}\label{bending_sheet5}
\alpha\approx
4\Bigl(\gamma\frac{\Dox}{\e}\Bigr)^2\frac{\rho_M^{1/2}(1+\nu)^{3/2}
 (1-\nu)^{1/2}}{\hbar t^2E^{3/2}(9+\frac{3\nu}{1-2\nu})}
\end{equation}
Here $\nu$ is the Poisson ratio of graphene. Choosing fairly
symmetrical TLSs, $\Dox/\e\sim1$, for the parameters in table
\ref{table_parameters}, $\alpha\sim10^{-5}\times (a/d)^{2n}$, very
small. In ref.\cite{E98} an expression is given for the inverse
quality factor of a vibration damped by TLSs in amorphous
insulators,
\begin{equation}\label{Qgeneral}
    Q^{-1}(\om,T)=\frac{P\gamma^2}{EkT}\int_0^{\e_{max}} d\e\int_{u_{min}}^{1}
du\frac{\om }{u\sqrt{1-u^2}}\,C(\om,T)
\end{equation}
where $u=\Dr/\e$, $\e_{max}\sim5$ K, and $(u\sqrt{1-u^2})^{-1}$
comes from the probability density of TLS's in an amorphous solid,
like SiO$_2$. $Q^{-1}(\om,T)$ is a function of $C(\om,T)$, the
Fourier transform of the correlation function
$C(t,T)=\langle\sigma_z(t)\sigma_z(0)\rangle_T$. For biased TLSs and
$\alpha\ll1$ an extensive analysis of $C(\om,T)$ is performed
in\cite{W99}, where several expressions are provided in different
limits. Using them, the estimate for $Q^{-1}(\om,T)$ follows:
\begin{equation}\label{limitsTLSs}
    \left\{
      \begin{array}{ll}
        Q^{-1}(\om,T)\approx\frac{P\gamma^2}{E\hbar\om}\Bigl\{\frac{4\pi}{3}\alpha\e_{max}+\frac{\pi^2}{3}\alpha^2kT\Bigr\}, & \hbox{$kT>\e_{max}$} \\
        Q^{-1}(\om,T)\approx\frac{P\gamma^2\alpha}{E\hbar\om}\frac{4\pi}{3}kT, & \hbox{$kT<\e_{max}$}
      \end{array}
    \right.
\end{equation}

In the range of temperatures of current experiments (5K$<$T$<$300K),
the dependence of dissipation with T is weak, and $Q^{-1}\sim
10^{-6}\times (a/d)^4\sim 10^{-22}$. The main uncertainty of the
calculation has been the use of the TLSs' distribution assumed for
amorphous solids \cite{P87}, but due to the small value of $\alpha$
a weak dissipation is expected also with a modified distribution.
Thus the conclusion is that the relative importance of TLSs damping
is much smaller for graphene than for other NEMs
devices\cite{CR02,B04,ER05}.

\section{Other friction mechanisms}
\subsection{Attachment losses.}
The energy is transferred from the
resonator mode to acoustic modes at the contacts and
beyond\cite{JI68,PJ04}.

The main expressions needed are given in\cite{PJ04}. When $d\gg t$,
and $d$ is much smaller than the wavelength of the radiated elastic
waves in the SiO$_2$ substrate, the contribution to the inverse
quality factor is given by
\begin{equation}\label{atloss1}
    Q^{-1}\approx \frac{w}{L}\Bigl(\frac{t}{d}\Bigr)^2
    \sqrt{\frac{\rho_M^CE^C(1-(\nu^O)^2)}{\rho_M^OE^O}}
\end{equation}
where the superscript $O$ applies to the silicon oxide, and $\nu^O$
stands for Poisson's ratio. The range of values of the quality
factor varies from $Q^{-1}\approx5\cdot10^{-6}$ for a graphene
monolayer, to $Q^{-1}\approx5\cdot10^{-3}$ for a stack with 30
layers and $t = 10$nm. These quantities probably overestimate the
attachment losses, as they do not include the impedance at the
SiO$_2$-graphene interface.

This damping process due to energy irradiated away from the
resonator should not depend on temperature.

\subsection{Thermoelastic effects.} When the phonon mean free path of
the acoustic phonons is shorter than the wavelength of the mode
under study, the acoustic phonons can be considered a dissipative
environment coupled to the mode by anharmonic terms in the ionic
potential\cite{Z38,LR00,U06}. These anharmonic effects are described
by the expansion coefficient, $\alpha$, and the thermal
conductivity, $\kappa$. We follow the analysis in\cite{Z48}. For a
rectangular beam vibrating at a frequency $\omega$ the inverse
quality factor is
\begin{equation}\label{Zenerdiss}
    Q^{-1}_Z(T)=\frac{E\alpha^2T}{C_p}\frac{\omega\tau_Z}{1+(\omega\tau_Z)^2}
\end{equation}
where $E$ is the Young Modulus, $C_p$ is the specific heat at a
constant pressure, and $\tau_Z$ is the thermal relaxation time
associated with the mode, which in the case of a flexural vibration
is given by $\tau_Z=t^2C_p/(\pi^2\kappa)$. This estimate assumes
that the graphene sheet is weakly deformed, and that the typical
relaxation time is associated to the diffusion of phonons over
distances comparable to the thickness of the sheet.

Although better approximations are available in the literature\cite{LR00},
eq.(\ref{Zenerdiss}) is enough for an estimate of the order of
magnitude of $Q^{-1}$.
Using the parameters from table[\ref{table_parameters}], for $t=10$
nm and $f\sim100$ MHz, we find that
 $\omega\tau_Z\ll 1$, and
\begin{equation}\label{Zenerdiss2}
    Q^{-1}_Z(T=300 \text{K})\approx\frac{E\alpha^2T\omega
t^2}{\pi^2\kappa}\sim5\cdot10^{-7}
\end{equation}

\section{Extension to nanotube oscillators.}
The analysis presented here can be extended, in a straightforward
way, to systems where the oscillating part is a nanotube.

We expect in these devices a larger
impedance between the modes of the nanotube and those of the
substrate, so that attachment losses will be suppressed with respect
to the estimate presented here for graphene.

The damping mechanisms which require long range forces between the
moving charges in the nanotube and degrees of freedom of the
substrate (fluctuating and static charges) will not be significantly
changed. A nanotube of length $L$ at distance $d$ from the substrate
will interact with a substrate area of order $( L + d ) \times d$. A
similar estimate for a graphene sheet of length $L$ and width $w$
gives an area $\sim ( L + d ) \times ( w + d )$. As $L \sim w \sim d
\sim 1 \mu$m, the two areas are comparable.

On the other hand,
ohmic losses induced in the nanotube will be reduced with respect to
the two dimensional graphene sheet, as the number of carriers is
lower in the nanotube.

 Finally, we expect a longer
phonon mean free path in the nanotube, which implies that
thermoelastic effects will be reduced .

\begin{table}[!t]
\begin{center}
\begin{tabular}{||l|c|c||}
\hline \hline
& $Q^{-1} ( T = 300 K )$ &Temperature \\ & &dependence \\
\hline Charges in the SiO$_2$ &$10^{-7} - 10^{-6}$ &$T$ \\
  Charges in graphene sheet &$10^{-2}$ &$T$ \\
 and metallic gate & & \\
 Velcro effect &\rm{Absent} &$T^0$ \\
  Two-level systems &$10^{-22}$ &$A+BT$ \\
 Attachment losses& $10^{-6} - 10^{-5}$ & $T^0$ \\
 Thermoelastic losses &$10^{-7}$ &$T$
\\ \hline \hline
\end{tabular}
\end{center}
\caption{Contribution of the mechanisms considered in the main text
to the inverse quality factor $Q^{-1}(T)$ of the systems studied
in\protect{\cite{Betal07}}.} \label{table_results}
\end{table}

\section{Conclusions.}
We have considered six possible dissipation mechanisms which may
lead to damping in a graphene mesoscopic oscillator. The main
results are summarized in Table[\ref{table_results}]. We expect that
the calculations give the correct order of magnitude and dependence
on external parameters.

We find that at high temperatures the leading damping mechanism is
the ohmic losses in the metallic gate and the graphene sheet. This
effect depends quadratically with the total charge at the graphene
sheet, which can be controlled by the gate voltage.

At low temperatures attachment losses limit the quality of the
vibration. If the resonator is strongly driven, a new damping
mechanism may come into play, the Velcro effect, which may limit
substantially the quality factor as compared with the slightly
driven case. The high crystallinity of the resonators eliminates the
main source of dissipation in semiconducting resonators, namely
surface-related effective TLSs coupled to the local strain field.

These conclusions apply with only slight modifications to carbon
nanotube-based resonators.

\section{Acknowledgements.}
F. G. acknowledges funding from MEC (Spain) through grant
FIS2005-05478-C02-01 the European Union Contract 12881 (NEST), and
CAM (Madrid), through program CITECNOMIK. A. H. C. N. was supported
through NSF grant DMR-0343790. We acknowledge many useful
discussions with A. M. Van Der Zande and J. Bunch.

\appendix
\section{Coupling to fixed charges in the ${\rm SiO}_2$ substrate}
\label{app_charge} The Fourier transform of the potential in
eq.(\ref{potential_charge})is:
\begin{equation}\label{potential_transform}
V ( {\bf \vec{q}} , \omega ) =2\pi e^2A e^{-qd}\delta (\omega - \wo)
\end{equation}
This potential is screened by the polarizability of the graphene
layer\cite{WSSG06}, so that $e^2$ has to be replaced by :
\begin{equation}\label{escreened}
e^2 \rightarrow {e^*}^2 = \frac{e^2}{1 + e^2 / | {\bf \vec{q}} |
{\rm Re} [\chi_0 ( | {\bf\vec{q}} | , \omega ) ]} \approx \frac{|
{\bf \vec{q}} |}{{\rm Re} [\chi_0 ( | {\bf \vec{q}} | , \omega ) ]}
\end{equation}
where $\chi_0$ is the susceptibility of the graphene layer. At low
energy and momenta its value tends to the compressibility of the
electrons in the layer:
\begin{equation}\label{compressgraph}
\lim_{| {\bf \vec{q}} | \rightarrow 0 , \omega \rightarrow 0} {\rm
Re} [ \chi_0 ( | {\bf\vec{q}} | , \omega ) ] = \left\{
\begin{array}{lr} \frac{\kf}{\vf}
&N=1 \\ \frac{N \gamma}{\vf^2} &N \ne 1 \end{array} \right.
\end{equation}
where $N$ is the number of layers and $\gamma$ is the interlayer
hopping element. For a stack with $N$ layers, we have used the model
with one interlayer hopping element\cite{GNP06}, which gives rise to
$2 N$ low energy bands, most of which show a quadratic dispersion.

Using Fermi's golden rule, we finally find for width of the graphene
mode ($v({\bf \vec{q}},\om)$=$v(-{\bf \vec{q}},\om)$):
\begin{equation}\label{FGRschematic}
\Gamma_{\rm ph} \approx \int d^2 {\bf \vec{q}} | v ( {\bf \vec{q}} )
|^2 {\rm Im} \chi_0 ( {\bf \vec{q}} , \wo ) \label{Fermi_charge}
\end{equation}
where:
\begin{equation}\label{Imsusc1}
{\rm Im} \chi_0 ( {\bf \vec{q}} , \wo ) \approx \left\{
\begin{array}{lr}
    \frac{| \omega | \kf}{\vf^2 | {\bf \vec{q}} |} &N=1 \\
\frac{| \omega | \gamma^2 N^{3/2}}{\vf^2 | {\bf \vec{q}} |
\sqrt{\rho}} &N \ne 1
    \end{array} \right.
\label{susc_charge}
\end{equation}
where, for $N \ne 1$, $\rho$ is the total carrier density. This last
expressions are valid for lengths bigger than the mean free path,
$l\gg l_{mfp}$.

The energy absorbed per cycle of oscillation and unit volume will be
$\Delta E=(2\pi/\wo)\hbar\wo\Gamma_{\rm ph}/twL=2\pi\hbar\Gamma_{\rm
ph}/twL$, and the inverse quality factor $Q^{-1}_{\rm ph}(\wo)$ will
correspond to
\begin{equation}\label{Qfactor1a}
    Q^{-1}_{\rm ph}(\wo)=\frac{1}{2\pi}\frac{\Delta E}{E_0}=\frac{\hbar\Gamma_{\rm ph}}{twL}\frac{1}{\frac{1}{2}\rho\wo^2A^2}=
    \frac{2\hbar\Gamma_{\rm ph}}{M\wo^2A^2}\,\,,
\end{equation}
where $E_0$ is the elastic energy stored in the vibration, $M$ is
the total mass of the resonator, and $A$ the amplitude of vibration.
Substituting
eqs.(\ref{potential_transform},\ref{escreened},\ref{compressgraph},\ref{Imsusc1})
in eq.(\ref{FGRschematic}), and inserting (\ref{FGRschematic}) in
eq.(\ref{Qfactor1a}), one arrives at eqs.(\ref{charge}) and
(\ref{charge_N}) for the dissipation due to a single charge in the
substrate. The analysis presented here does not consider additional
screening due to the presence of a metallic gate. In that case, one
needs to add to the potential from a static charge,
eq.(\ref{potential_charge}) in the main text, a contribution from
the image charge induced by the gate. This effect will reduce the
coupling between the graphene layer and charges in the vicinity of
the gate.

\section{Screening of the potentials at the graphene sheet and Si gate}\label{metallic_layers}
The equations for the selfconsistent potentials $v_{scr}(z,{\bf
\vec{r}}-{\bf \vec{r}}',\om)$ as a function of the bare potentials
$v^j_0(z,{\bf \vec{r}}-{\bf \vec{r}}',\om)$ are given by

\begin{widetext}
\begin{eqnarray}
  \nonumber v_{scr}(d,{\bf \vec{r}}-{\bf \vec{r}}',\om) &=& v^C_0(d,{\bf \vec{r}}-{\bf \vec{r}}',\om)+v^G_0(d,{\bf \vec{r}}-{\bf \vec{r}}',\om)
  +\int_Cd{\bf \vec{r}}_1\int_Cd{\bf \vec{r}}_2
   v_{\text{Coul}}(d,{\bf \vec{r}}-{\bf \vec{r}}_1,\om)\cg({\bf \vec{r}}_1-{\bf \vec{r}}_2,\om)v_{scr}(d,{\bf \vec{r}}_2-{\bf \vec{r}}',\om)+\\
  \nonumber &+& \int_Gd{\bf \vec{r}}_3\int_Gd{\bf \vec{r}}_4
   v_{\text{Coul}}(d,{\bf \vec{r}}-{\bf \vec{r}}_3,\om)\csi({\bf \vec{r}}_3-{\bf \vec{r}}_4,\om)v_{scr}(0,{\bf \vec{r}}_4-{\bf \vec{r}}',\om) \\
  \nonumber v_{scr}(0,{\bf \vec{r}}-{\bf \vec{r}}',\om) &=& v^G_0(0,{\bf \vec{r}}-{\bf \vec{r}}',\om)+v^C_0(0,{\bf \vec{r}}-{\bf \vec{r}}',\om)+
  \int_Gd{\bf \vec{r}}_1\int_Gd{\bf \vec{r}}_2
   v_{\text{Coul}}(0,{\bf \vec{r}}-{\bf \vec{r}}_1,\om)\csi({\bf \vec{r}}_1-{\bf \vec{r}}_2,\om)v_{scr}(0,{\bf \vec{r}}_2-{\bf \vec{r}}',\om)+ \\
   &+& \int_Cd{\bf \vec{r}}_3\int_Cd{\bf \vec{r}}_4
   v_{\text{Coul}}(0,{\bf \vec{r}}-{\bf \vec{r}}_3,\om)\cg({\bf \vec{r}}_3-{\bf \vec{r}}_4,\om)v_{scr}(d,{\bf \vec{r}}_4-{\bf
   \vec{r}}',\om)\,\,,
\end{eqnarray}
\end{widetext}

where for example in the first equation $v^C_0(d,{\bf \vec{r}}-{\bf
\vec{r}}',\om)$ represents the bare potential experienced by a point
charge $e$ in the graphene layer due to the presence of charges in
that same layer, while $v^G_0(d,{\bf \vec{r}}-{\bf \vec{r}}',\om)$
is the bare potential experienced by a point charge $e$ in the
graphene layer due to the presence of charges in the Si plane.
$v_{\text{Coul}}$ is the two-dimensional bare Coulomb potential.
These equations simplify considerably in the ${\bf \vec{q}}$ space:

\begin{widetext}
\begin{equation}\label{scr1}
    \left\{
      \begin{array}{l}
         v_{scr}(d,{\bf \vec{q}},\om)=v^C_0(d,{\bf \vec{q}},\om)e^{qd}+ v^G_0(d,{\bf \vec{q}},\om)+v_q\cgq v_{scr}(d,{\bf \vec{q}},\om)+
  v_qe^{-qd}\csiq v_{scr}(0,{\bf \vec{q}},\om)\\
\\
         v_{scr}(0,{\bf \vec{q}},\om)=v^C_0(0,{\bf \vec{q}},\om)+ v^G_0(d,{\bf \vec{q}},\om)e^{qd}+v_qe^{-qd}\cgq v_{scr}(d,{\bf \vec{q}},\om)+
  v_q\csiq v_{scr}(0,{\bf \vec{q}},\om)
      \end{array}
    \right.\,\,,
\end{equation}
\end{widetext}

where $v_q=2\pi e^2/|{\bf \vec{q}}|$ is the Fourier transform of the
Coulomb potential in two dimensions, and where $v^G_0(0,{\bf
\vec{q}},\om)$ and $v^C_0(d,{\bf \vec{q}},\om)$ have been expressed
in terms of $v^G_0(d,{\bf \vec{q}},\om)$ and $v^C_0(0,{\bf
\vec{q}},\om)$. Now we can calculate $v_{scr}(d,{\bf \vec{q}},\om)$
and $v_{scr}(0,{\bf \vec{q}},\om)$ in terms of the rest of the
variables,

\begin{widetext}
\begin{equation}
  \left(
      \begin{array}{c}
         v_{scr}(d)\\
        v_{scr}(0) \\
      \end{array}
    \right)= \left(
              \begin{array}{cc}
                1-v_q\cg & -v_qe^{-qd}\csi \\
                -v_qe^{-qd}\cg & 1-v_q\csi \\
              \end{array}
            \right)^{-1}\times \left(
                             \begin{array}{cc}
                                e^{qd} & 1 \\
                               1 & e^{qd} \\
                               \end{array}
                           \right)\left(
                     \begin{array}{c}
                       v_0^C(0) \\
                       v_0^G(d) \\
                     \end{array}
                   \right) \\
\end{equation}
\end{widetext}

The dependence on ${\bf \vec{q}}$ and $\om$ has been omitted for the
sake of clarity. Now, if we are interested only in the long
wavelenght limit $v_q\cg,v_q\csi\gg 1$, the last equation simplifies
to

\begin{widetext}
\begin{equation}\label{vscr2}
    \left(
      \begin{array}{c}
         v_{scr}(d)\\
        v_{scr}(0) \\
      \end{array}
    \right)=\frac{1}{v_q^2\cg\csi\Bigl(1-e^{-2qd}\Bigr)}\times\left(
    \begin{array}{cc}   v_q\Bigl(\cg e^{-qd}-\csi e^{qd}\Bigr) &
      v_q\Bigl(-\csi+\cg\Bigr) \\  v_q\Bigl(-\cg+\csi\Bigr) &
      v_q\Bigl(\csi e^{-qd}-\cg e^{qd}\Bigr) \\   \end{array}  \right)\left(
                     \begin{array}{c}
                       v_0^C(0) \\
                       v_0^G(d) \\
                     \end{array}
                   \right)
\end{equation}
\end{widetext}

\subsubsection{Values of $v_0^C(0,{\bf \vec{q}},\om)$ and $v_0^G(d,{\bf \vec{q}},\om)$}
Now we will calculate the parts of these terms which will give rise
to a coupling to the vibration. When the graphene layer is set into
motion with a bending mode of wavevector ${\bf \vec{q}}$ and
amplitude $A_{\bf \vec{q}}$, the potential of a point charge $e$ in
the Si plane due to the charge in the graphene layer, $v_0^C(0,{\bf
\vec{r}},t)$, is

\begin{widetext}
\begin{eqnarray}\label{voz}
    \nonumber v_0^C(0,{\bf \vec{r}},t)&=& \frac{1}{2}\int_Cd{\bf \vec{r}}'v_{\text{Coul}}({\bf \vec{r}}-{\bf \vec{r}}',z')\rho({\bf
 \vec{r}}',z',t)=\frac{1}{2}\int_Cd{\bf \vec{r}}'\frac{2\pi e^2\rho_0}{\sqrt{({\bf \vec{r}}-{\bf \vec{r}}')^2 +(d+A_{\bf \vec{q}}e^{i({\bf \vec{q}}
 {\bf \vec{r}}'-\wq t)})^2}}\\&\approx&\frac{1}{2}\int_Cd{\bf \vec{r}}'\frac{2\pi e^2\rho_0}{\sqrt{({\bf \vec{r}}-{\bf
    \vec{r}}')^2+d^2}}+\frac{1}{2}\int_Cd{\bf \vec{r}}'\frac{2\pi e^2\rho_0A_{\bf \vec{q}}e^{i({\bf \vec{q}}{\bf \vec{r}}'-\wq t)}d}
    {\Bigl(({\bf \vec{r}}-{\bf \vec{r}}')^2+d^2\Bigr)^{3/2}}\approx
    f({\bf \vec{r}})+\pi e^2\rho_0A_{\bf \vec{q}}e^{-dq}e^{i({\bf \vec{q}}{\bf \vec{r}}-\wq t)}
\end{eqnarray}
\end{widetext}

where in the second line an expansion for small $A_{\bf \vec{q}}$
has been performed. The Fourier transform for $\om\neq0$ is
\begin{equation}\label{voz2}
    v_0^C(0,{\bf \vec{k}},\om')=\pi e^2\rho_0A_{\bf
    \vec{q}}e^{-dq}\delta({\bf \vec{k}}-{\bf
    \vec{q}})\delta(\om'-\wq)\,\,\,\,,\,\,\,|{\bf \vec{q}}|=1/L
\end{equation}

Similarly, the potential of a point charge in the oscillating
graphene sheet due to the charge in the Si plane $v_0^G(d)$, is

\begin{widetext}
\begin{equation}\label{voz3b}
    v_0^G(d,{\bf \vec{r}},t)= \frac{1}{2}\int_Gd{\bf \vec{r}}'\frac{2\pi e^2\rho_0}
 {\sqrt{({\bf \vec{r}}-{\bf \vec{r}}')^2 +(d+A_{\bf \vec{q}}e^{i({\bf \vec{q}}{\bf \vec{r}}-\wq t)})^2}}\approx
  f({\bf \vec{r}})+\pi e^2\rho_0A_{\bf \vec{q}}e^{i({\bf \vec{q}}{\bf \vec{r}}-\wq t)}
\end{equation}
\end{widetext}

leading to the same expression as eq.(\ref{voz2}) but without the
factor $e^{-qd}$
\begin{equation}\label{voz4}
    v_0^G(d,{\bf \vec{k}},\om')=v_0^C(0,{\bf \vec{k}},\om')e^{qd}
\end{equation}
Substituting (\ref{voz2},\ref{voz4}) in eq.(\ref{vscr2}), one
obtains eq.(\ref{vscr0}).

\bibliography{NEMs1941}

\end{document}